\documentclass[pra,aps,twocolumn,groupedaddress,a4paper,showpacs,nofootinbib]{revtex4}

\usepackage{graphicx}
\usepackage{amsmath}
\usepackage{latexsym}
\usepackage{amsfonts}
\usepackage{amssymb}
\usepackage{amsthm}
\usepackage{makeidx}
\usepackage{bm}
\usepackage{verbatim}
\usepackage{lscape}

    \newcommand{\bra}[2][]{#1\langle #2 #1\rvert}
    \newcommand{\ket}[2][]{#1\lvert #2 #1\rangle}
    \newcommand{\braket}[3][]{#1\langle #2 #1| #3 #1\rangle}
    
\newcommand{\pderiv}[2]{\ensuremath{\frac{\partial #1}{\partial #2}}}

   \newcommand{\tr}[0]{\mathrm{tr}}

\newcommand{\matrixel}[3]{\ensuremath{\langle #1 \lvert #2 \rvert #3 \rangle}}
\newcommand{\qmproj}[1]{\ensuremath{\ket{#1} \bra{#1}}}

\newtheorem{thm}{Theorem}
\newtheorem{lem}{Lemma}

\begin{document}

\title{Energy as an Entanglement Witness for Quantum Many-Body Systems}
\author{Mark R. Dowling}
\email{dowling@physics.uq.edu.au}
\author{Andrew C. Doherty}
\author{Stephen D. Bartlett} 
\affiliation{School of Physical Sciences, The University of Queensland,
  Queensland 4072, Australia}

\date{4 January 2005}

\begin{abstract}
  We investigate quantum many-body systems where all low-energy states
  are entangled.  As a tool for quantifying such systems, we introduce
  the concept of the \emph{entanglement gap}, which is the difference
  in energy between the ground-state energy and the minimum energy
  that a separable (unentangled) state may attain.  If the energy of
  the system lies within the entanglement gap, the state of the system
  is guaranteed to be entangled.  We find Hamiltonians that have the
  largest possible entanglement gap; for a system consisting of two
  interacting spin-$1/2$ subsystems, the Heisenberg antiferromagnet is
  one such example.  We also introduce a related concept, the
  \emph{entanglement-gap temperature}: the temperature below which the
  thermal state is certainly entangled, as witnessed by its energy.
  We give an example of a bipartite Hamiltonian with an arbitrarily
  high entanglement-gap temperature for fixed total energy range.  For
  bipartite spin lattices we prove a theorem demonstrating that the
  entanglement gap necessarily decreases as the coordination number is
  increased.  We investigate frustrated lattices and quantum phase
  transitions as physical phenomena that affect the entanglement gap.
\end{abstract}


\maketitle

\section{Introduction}

Understanding and quantifying the properties of quantum many-body
systems is a central goal of theoretical condensed matter physics.
Progress is often hindered by an incomplete understanding of the
highly non-classical entangled states that occur naturally as the
ground and thermal states of many systems.  Entanglement is perhaps
the most counter-intuitive feature of quantum mechanics, and results
in stronger correlations than can be present in any classical system
\cite{EPR35,bell}.  Recently, entanglement has been recognized as an
important resource in the emerging field of quantum information
science~\cite{NC}, which has led to new tools that may enhance our
understanding of the role of entanglement in quantum many-body
systems.

Much recent work has focused on quantifying the entanglement naturally
present in the ground state of standard models of coupled quantum
systems, particularly spin chains.  In
\cite{ON02,OAF+02,VLR+03,latorre2003a,WDM+,somma2004a,lidar} the role of
entanglement in a \emph{quantum phase transition} \cite{Sac} is
investigated.  In one-dimensional chains, the amount of entanglement
between a length of spins and the rest of the chain appears to depend
only on the universality class of the model at the phase
transition~\cite{VLR+03,latorre2003a}. Various quantities associated
with entanglement have been shown to display universal scaling
behavior at phase transitions in one dimension
\cite{ON02,OAF+02,WDM+}.  Also, it appears that properties of
entanglement between spins, such as the entanglement length defined in
\cite{VPC04}, are sometimes able to characterize phases of the system
better than any correlation length~\cite{VMC04}.

Restricting to many-body systems where each system interacts with only
a finite local neighborhood (which we refer to as \emph{local}
interactions) very strongly constrains the quantum states that must be
considered.  For example, there exist quantum states that are far from
the ground state of any local-interaction Hamiltonian~\cite{HON02}.
For finite systems with local interactions and an energy gap it was
shown in \cite{haselgrove2004a} that there are strong bounds on both
the correlations and the entanglement in the ground state.  The fact
that local interactions strongly limit the entanglement that can occur
for Hamiltonian systems with local interactions on a line or a plane
has been used to develop new approximation schemes for simulating
quantum dynamics \cite{V03,V,DKS+,CJ,ZV,VGC}.  There is now a large
literature on the entanglement properties of the ground states of
Hamiltonian systems; we refer the reader to
\cite{oconnor2001a,costi2003a,hines2003a,tessier2003a,hines2003b,
  dawson2004a} and references therein.

Although the ground state plays an essential role in understanding
physical systems, at finite temperature it is the thermal state that
is of the greatest interest.  The nature of entanglement in the
thermal state of condensed matter systems was first studied by Nielsen
\cite{NPhD}, who investigated how entanglement in the thermal state
varied with temperature and other parameters of simple systems
consisting of two coupled spins.  Subsequent work has investigated
similar questions for quantum many-body systems
\cite{arnesen2001a,gunlycke2001a,wang2001a,scheel2003a,jordan2004,VedA,VedB}.
A recent experiment demonstrates that entanglement can affect
thermodynamic properties of a system at high temperature
\cite{GRA+03}.

Thus, it seems that many physical phenomena involving just the ground
or thermal states in condensed matter systems may be associated with
the nature of entanglement in the system, and it is important to
investigate new techniques for understanding and quantifying the role
of entanglement in such systems.  Desirable features of these
techniques include that they be easily computable, even for large
systems, that they be applicable at finite temperature and that they
be in principle easy to measure and related to known physics.  Because
most quantum systems are described by mixed states, these criteria lead
naturally to the theory of mixed state entanglement.

Surprisingly, even the question of whether a mixed state of a quantum
system is entangled or not is a difficult and much studied question.
We refer the reader to the many reviews for the literature on the
so-called separability problem
\cite{terhal2002a,lewenstein2000b,bruss2002a}. The difficulty of this
problem is one of the reasons why computing measures of entanglement
can be so challenging and why it is important to find more tractable
ways of understanding the entanglement in real physical systems. In
this work apply results from the theory of the separability
problem to study the entanglement of quantum many-body systems. This
investigation leads both to an understanding of the kind of
Hamiltonians possessing strongly entangled thermal and low energy
states and also to interesting connections with properties of spins
systems studied in more conventional condensed matter approaches.

The specific concept that we use from the theory of separability is
that of an \emph{entanglement witness}: an observable whose
expectation value is positive if the state of interest is not
entangled but for which a negative expectation value indicates that
the state is entangled. An example of such an observable is the Bell
observable which describes the outcomes of a test for the violation of
Bell inequalities. In this paper we explore the idea of interpreting
Hamiltonians with entangled ground states as entanglement witnesses.
This point of view has attracted interest recently.  During the
preparation of this paper, related investigations appeared by Brukner
and Vedral \cite{BV} and by T\'oth \cite{TC} in which this type of
entanglement witness is studied.  As emphasized by Brukner and Vedral
\cite{BV}, because energy is a macroscopic thermodynamic property it
is reasonable to expect that it could be measured in experiment.

In this paper, we develop the idea of using energy as an entanglement
witness for quantum many-body systems.  We introduce two related
concepts inspired by the theory of entanglement witnesses, and discuss
their relevance to both ground-state and finite-temperature properties
of quantum many-body systems.  The first is the \emph{entanglement
  gap}: the difference in energy between the ground-state energy and
the minimum energy that any separable (unentangled) state may attain.
If the entanglement gap of a system is non-zero, then the entanglement of
certain mixed states is detected simply by measuring their energy to
be below this threshold.  Roughly speaking, if the entanglement gap is
small then a separable state can be a good approximation to the ground
state, and we expect approximation schemes based on separable states
to produce reliable results. We investigate how large this gap can be
for two-spin systems and how this gap depends on the coordination
number for lattices of coupled spins.

One advantage of using ideas from studies of mixed-state entanglement is that it is possible to investigate systems at
finite temperature.  The second concept we introduce is the use of
temperature as an indicator of entanglement in the thermal state.  By
comparing the thermal energy with the entanglement gap we obtain a
temperature threshold, the \emph{entanglement-gap temperature}, below
which the thermal state is certainly entangled and we may expect
entanglement to influence thermodynamic properties. We show that this
temperature can become arbitrarily large as the dimension of two
interacting spins increases even if the energy range of the system is
kept fixed.

We begin in Sec.~\ref{sec: hamentwit} with the observation that
Hamiltonians with entangled ground states may be viewed as
entanglement witnesses.  We introduce the notion of entanglement gap
and provide necessary and sufficient conditions for this gap to be
non-zero. We construct a one-to-one mapping between Hamiltonians with
non-zero entanglement gap and entanglement witnesses.  We prove a
theorem that identifies a set of Hamiltonians with the largest
possible gap; for spin-1/2 particles, one such Hamiltonian is the
Heisenberg antiferromagnet.  We then formally define the
entanglement-gap temperature and investigate conditions that lead to a
high value of this temperature.  Somewhat counterintuitively, the
Hamiltonians with the largest entanglement gap do not have the largest
entanglement-gap temperature in general.

In Sec.~\ref{sec: multipartite} we study the entanglement gap in
many-body systems, in particular spin models on lattices.  We prove a
general result that the entanglement gap must go to zero with
increasing coordination number on a bipartite lattice with a fixed
local interaction.  This result is suggestive of a relationship to the
success of mean-field theory on lattices with high coordination
number.  In Sec.~\ref{sec:Discussion}, we conclude by investigating
the dependence of the entanglement gap on frustration for the
Heisenberg antiferromagnet.  We show that for such systems it is
possible to determine that the system is entangled even when the
reduced state of nearest neighbor spins is not entangled.  We also
investigate how the entanglement gap behaves near the quantum phase
transition in the $XY$ model and discuss its relationship to previous
studies of entanglement at this transition.

\section{Hamiltonians as Entanglement Witnesses}
\label{sec: hamentwit}

In this section we establish a formal connection between Hamiltonians
with the property that all low-energy states up to a certain energy
are entangled, and entanglement witnesses.

A multipartite mixed state of $n$ subsystems is said to be
\emph{separable} if it can be expressed as a convex combination of
pure product states
\begin{equation}
  \rho = \sum_{i} p_i \qmproj{\psi^1_i} \otimes
  \qmproj{\psi^2_i} \otimes \ldots \otimes  \qmproj{\psi^n_i} \,, 
\end{equation}
where $\ket{\psi^j_i}$ are pure states in the Hilbert space
$\mathcal{H}_j$ of subsystem $j$, and $p_i > 0$, $\sum_i p_i = 1$.  If
a state can be decomposed in this way, then all correlations are
purely classical; if not, then there exist truly quantum correlations
and we say that the state is \emph{entangled}.

An entanglement witness, $Z_{EW}$, on a multipartite system is a
Hermitian operator (observable) with the properties that its expectation
value in any separable state is greater than or equal to zero
\begin{equation}
  \tr[Z_{EW} \rho_{\mathrm{sep}}] \geq 0\,, \quad \forall\
  \rho_{\mathrm{sep}} \in \mathcal{S}\,, 
\end{equation}
where $\mathcal{S}$ is the set of all separable states, and that there
exists an entangled state, $\rho_{\mathrm{ent}}$, such that
\begin{equation}
  \tr[Z_{EW} \rho_{\mathrm{ent}}] < 0 \,.
\end{equation}
We say that $Z_{EW}$ witnesses the entanglement of
$\rho_{\mathrm{ent}}$.

For a multipartite Hamiltonian, $H$, we define the \emph{minimum
  separable energy},
\begin{equation}
  E_{\mathrm{sep}} = \min_{\rho_{\mathrm{sep}} \in \mathcal{S}}  \tr
  [H \rho_{\mathrm{sep}}]\,.  
\end{equation}
Due to the convexity of the set of separable states, this minimum can
always be achieved by a pure separable state.  Note that there may be
many separable states achieving this minimum separable energy
$E_{\mathrm{sep}}$.

If $E_{\mathrm{sep}}$ is strictly greater than the ground-state
energy, $E_0$, there is a finite energy range over which all states
are entangled.  We refer to the size of this energy range as the
entanglement gap.

\medskip\noindent
\textbf{Definition:} For any multipartite Hamiltonian, $H$, we define
the \emph{entanglement gap},
\begin{equation}
  G_E = E_{\mathrm{sep}}  - E_0 \,,
\end{equation}
where $E_0$ is the ground-state energy of $H$.  The entanglement gap
is the energy gap between the ground-state energy and the minimum
energy that a separable state can attain. \medskip

If $H$ has an entangled non-degenerate ground state $\ket{E_0}$ then
any separable state written in terms of the eigenstates of the
Hamiltonian must contain contributions from higher energy states and
must therefore have higher energy than $E_0$.  If the ground state is
degenerate the same argument requires that the entanglement gap is
greater than zero if there is no state in the ground-state manifold
that is a pure product state.  Conversely a non-zero entanglement gap
requires that $E_{\mathrm{sep}}>E_0$ and so there can be no pure
product state in the ground-state manifold because such a state would
have energy $E_0$.  So whether or not the entanglement gap is zero
depends only on the ground-state manifold. \emph{A Hamiltonian $H$ has
  a non-zero entanglement gap if and only if no ground state of $H$ is
  separable.}

Constructing Hamiltonians with a non-zero entanglement gap is
straightforward.  Every entanglement witness can be regarded as a
Hamiltonian for a multipartite quantum system.  For such Hamiltonians,
$E_{\mathrm{sep}}=0$ and $E_0$ is the minimum eigenvalue of $Z_{EW}$.
The definition of entanglement witnesses implies that $E_0<0$ and thus
the entanglement gap is non-zero.

Every Hamiltonian with a positive entanglement gap $G_E>0$ defines an
entanglement witness,
\begin{equation}
  Z_{EW} = H -  E_{\mathrm{sep}} I \,,
\end{equation}
where $I$ is the identity on the total Hilbert space.  Because
$E_{\mathrm{sep}}$ is the lowest possible energy for a separable state
we have $\tr[Z_{EW} \rho_{\mathrm{sep}}]=\tr[H
\rho_{\mathrm{sep}}]-E_{\mathrm{sep}} \geq 0$. On the other hand if
$\rho_0$ is a state in the ground-state manifold we have $\tr[Z_{EW}
\rho_{\mathrm{sep}}]= E_0 -E_{\mathrm{sep}} < 0$ so $Z_{EW}$ is an
entanglement witness. Note that if $H'$ and $H$ differ only by an
additive constant they lead to the same entanglement witness. We
regard such Hamiltonians as equivalent.

In summary, \emph{there is a one-to-one map between entanglement
  witnesses and the equivalence classes of Hamiltonians with non-zero
  entanglement gap. }

The entanglement gap quantifies the range of energies over which all
states are necessarily entangled.  Note, however, that higher energy
states may still be entangled.  So, for example, the thermal state for
$H$ must be entangled for all temperatures such that the thermal
energy is below $E_{\mathrm{sep}}$ but at higher temperatures the
thermal state may or may not be entangled.

In appendix~\ref{app: semidefentgap} we describe an efficient numerical procedure, a sequence of semidefinite programs, for evaluating the entanglement gap and discuss the concept of bound entanglement in this context.

\subsection{Hamiltonians that Maximize the Entanglement Gap}

\label{subsec: bientgap}

Having defined the entanglement gap it is natural to identify
Hamiltonians that have the largest possible entanglement gap for a
given multipartite quantum system.  We proceed by proving two lemmas,
one that the entanglement gap is invariant under local unitary
transformations of the Hamiltonian, and the other regarding the
optimal arrangement of the energy levels.  We use these two lemmas to
prove the main theorem of this section, which is that a set of
Hamiltonians with maximum possible entanglement gap are those with a
non-degenerate maximally entangled ground state and all other
eigenstates at equal energy.

\begin{lem}
\label{lem: locunitary}
Given a multipartite Hamiltonian, $H$, and a local unitary $U_{\rm
  local} = U_1 \otimes U_2 \otimes\cdots\otimes U_N$ acting on each
subsystem, the Hamiltonian $H' = U_{{\rm local}} H U_{\rm
  local}^\dagger$ has the same entanglement gap as $H$.
\end{lem}

\noindent
\textbf{Proof:} From the cyclic property of the trace, we have $\tr[H'
\rho'_{\mathrm{sep}}] = \tr [H \rho_{\mathrm{sep}}]$, where
$\rho_{\mathrm{sep}} = U_{\rm local}^\dagger \rho'_{\mathrm{sep}}
U_{\rm local}$ is also separable.  That is, for each
$\rho_{\mathrm{sep}}$ with a certain energy under $H$ there is a
separable state $\rho'_{\mathrm{sep}}$ with the same energy under
$H'$.  Therefore $E_{\mathrm{sep}} = E'_{\mathrm{sep}}$.  Also,
because $H$ and $H'$ are related by conjugation by a unitary they have
the same spectrum, and in particular the same ground-state energy.
Hence $H$ and $H'$ have equal entanglement gap.  \hfill$\Box$\medskip

We now determine which Hamiltonians have the largest entanglement gap.
For a comparison of gaps to be a sensible, we need to scale by the
overall energy range of the system.  We define the \emph{scaled
  entanglement gap}, $g_E$ as
\begin{equation}
  g_E = G_E/E_{\mathrm{tot}}\,,
\end{equation}  
where $E_{\mathrm{tot}} = E_{\rm max} - E_0$ is the total energy range,
and $E_{\rm max}$ is the highest energy eigenvalue.

\begin{lem}
\label{lem: raiseintlevels}
For any Hamiltonian with scaled entanglement gap $g_E$, the
Hamiltonian $H' = I - \qmproj{E_0}$, where $\ket{E_0}$ is a ground
state for $H$, has a scaled entanglement gap $g'_E$ greater than or
equal to $g_E$.
\end{lem}

\noindent
\textbf{Proof:} We scale the Hamiltonian so that its lowest eigenvalue
is zero and its highest eigenvalue is one, and thus the energy
eigenvalues lie in the range $0 \leq E_i \leq 1$, $i = 0 \ldots d_T-1$, where $d_T$ is the dimension of the total Hilbert space.  The
entanglement gap $G_E$ of the scaled Hamiltonian $\bar H$ is equal to
the scaled entanglement gap $g_E$ of the original Hamiltonian $H$.
Note that the Hamiltonian $H'$ is already scaled in this manner, i.e.,
$g'_E = G'_E$.

To prove the lemma, it is sufficient to show the stronger result $\tr[\bar H \rho] \leq \tr[H' \rho], \quad \forall\ \rho$, i.e., all states have higher energy under $H'$ than under $\bar H$.  To this end
\begin{equation}
\tr[\rho \bar H] = \sum_{i=0}^{d_T-1} E_i \matrixel{E_i}{\rho}{E_i} \leq \sum_{i=1}^{d_T-1} \matrixel{E_i}{\rho}{E_i} = \tr [H' \rho] \,, 
\end{equation}
as required.  The inequality follows from the assumed range of
energies $0 \leq E_i \leq 1\ \forall\ i$ (where $E_0 =0$) and the fact
that $0 \leq \matrixel{\psi}{\rho}{\psi} \leq 1$ for any density
operator, $\rho$, and any state, $\ket{\psi}$ (because $0 \leq \rho
\leq I$).  Therefore $E'_{\rm sep}$ is necessarily greater than
$E_{\rm sep}$ (even if the minimum-energy separable states are
different), and because $E'_0 = E_0 = 0 $ and both Hamiltonians are
scaled appropriately, we have $g'_E \geq g_E$, as required.
\hfill$\Box$\medskip
  
Using the geometric measure of entanglement for multipartite systems
defined in \cite{wei2003a}, we consider multipartite pure states that
are maximally entangled in the sense that they have \emph{minimal}
overlap with any pure product state, i.e., that they maximize the
entanglement measure
\begin{equation}
  \label{eq:maxent}
  M(|\Psi\rangle) = 1 - \max_{\rho_{\mathrm{sep}\in \mathcal{S}}}
  \langle\Psi|\rho_{\mathrm{sep}}|\Psi\rangle \,. 
\end{equation}
Let $M^{\rm max} = M(|\Psi_{\rm me}\rangle)$ be the maximum value of
this measure, achievable by a maximally entangled state $|\Psi_{\rm
  me}\rangle$.

\begin{thm}
\label{thm: maxentgap}  
The largest possible scaled entanglement gap of a multipartite system is
$g_E^{\rm max} = M^{\rm max}$, and can be achieved by any Hamiltonian
of the form $H' = I - \qmproj{\Psi_{\rm me}}$, where $\ket{\Psi_{\rm
    me}}$ is a maximally entangled state by the measure of
Eq.~(\ref{eq:maxent}).
\end{thm}

\noindent
\textbf{Proof:} The proof follows from the definition of the
entanglement gap,
\begin{equation}
  g_E= 1 - \max_{\rho_{\mathrm{sep}\in \mathcal{S}}}
  \matrixel{E_0}{\rho_{\mathrm{sep}}}{E_0}\,,
\end{equation}
and from Lemma~\ref{lem: raiseintlevels}.
\hfill$\Box$\medskip

Although we do not present the result here, it is also possible to
show~\cite{harrow2004a} that all Hamiltonians that have this maximum
entanglement gap are of this form.

For multipartite systems it is not known which states are maximally
entangled according to the measure $M$.  However, in~\cite{wei2003a}
examples of highly entangled states are given, which place lower
bounds on the maximum size of the scaled entanglement gap.  For
example, if each of the $n$ subsystems are $d$-dimensional, there
exists a symmetrised state $\ket{S(n,d)}$ such that $M(\ket{S(n,d)})$
approaches $1$ as $d^{-2n}$ in the $n \to \infty$ limit.  If each of
the $n$ subsystems are $n$-dimensional, there is an antisymmetrised
state, $\ket{A(n)}$ such that $M(\ket{A(n)}) = 1 - 1/n!$.  It is clear
that entanglement gap can be a very large fraction of the total energy
range for large numbers of coupled systems.

Bipartite entanglement is much better understood than multipartite
entanglement, and the following Corollary gives an explicit form for the
maximally entangled ground state and the corresponding maximum possible
scaled entanglement gap for bipartite systems.

\medskip\noindent \textbf{Corollary:} \textit{The largest scaled
  entanglement gap for a bipartite system $\mathcal{H}_A \otimes
  \mathcal{H}_B$ is $g_E = 1 - 1/d$, where $d= \min(d_A,d_B)$ is the
  smaller dimension of the two subsystems, and is achieved by any
  Hamiltonian of the form $H' = I - \qmproj{\phi_d}$, where
  $\ket{\phi_d} = \frac{1}{\sqrt{d}} \sum_{i=1}^d \ket{i_A}
  \ket{i_B}$, and $\{ \ket{i_{A/B}} \}$, are orthonormal bases for
  $\mathcal{H}_{A/B}$.}
  
\medskip\noindent \textbf{Proof:} It follows from the convexity of the
set of separable density matrices that the maximum overlap between a
pure ground state and a separable state is achieved by a \emph{pure}
product state $\rho_{\mathrm{sep}} = \qmproj{A} \otimes \qmproj{B}$,
where $\ket{A} \in \mathcal{H}_A$, $\ket{B} \in \mathcal{H}_B$.  In
fact, the maximum is achieved by setting $\ket{A} = \ket{1_A}$,
$\ket{B} = \ket{1_B}$,
\begin{equation}
  \max_{\rho_{\mathrm{sep}\in \mathcal{S}}}
  \matrixel{E_0}{\rho_{\mathrm{sep}}}{E_0} = \lambda_1^2\,, 
\end{equation} 
where the Schmidt decomposition \cite{NC} for $\ket{E_0}$ is
$\ket{E_0} = \sum_{i = 1}^d \lambda_i \ket{i_A} \ket{i_B},$ and
$\lambda_1$ is the largest Schmidt coefficient.

Thus, the largest scaled entanglement gap results from finding
$\ket{E_0}$ with the smallest possible $\lambda_1$.  Normalization
($\sum_i \lambda_i^2 = 1$) requires that $\lambda_1^2 \geq 1/d$ and
$\lambda_1^2=1/d$ is achieved by any maximally entangled bipartite
state $\ket{E_0} = \ket{\phi_d}$.  Thus, the Hamiltonian
$H=I-\qmproj{\phi_d}$ achieves the maximum possible scaled
entanglement gap, $g_E = 1-1/d$.  \hfill$\Box$\medskip

For $d_A = d_B =2$, the Hamiltonian $H = I - \qmproj{\phi_2}$, where
$\ket{\phi_2} = \sum_{i=1}^2 \ket{i_A} \ket{i_B}/\sqrt{2}$ is any
maximally entangled state, has the largest entanglement gap.  If the
Hilbert space corresponds physically to two spin-$1/2$ systems, then a
particularly enlightening example of a Hamiltonian of this form is a
shifted and scaled version of the antiferromagnetic Hamiltonian, $H =
\vec{\sigma}_A \cdot \vec{\sigma}_B$, where $\vec{\sigma}_i$, $i =
A,B$ is the vector of Pauli matrices on $\mathcal{H}_i$.  It is
straightforward to show that
\begin{equation*}
  H = I - \qmproj{\psi^-} = (\vec{\sigma}_A \cdot \vec{\sigma}_B + 3
  I)/4 \,,
\end{equation*}
where $\ket{\psi^-} = (\ket{0}_A \ket{1}_B - \ket{1}_A \ket{0}_B
)/\sqrt{2}$ is the singlet state.

\subsection{Entanglement-Gap Temperature}

\label{subsec: bienttemp}

In the following, we investigate the temperature at which the thermal
state reaches the minimum separable energy.  We find the temperature
below which the thermal state is guaranteed to be entangled; this
temperature also provides a non-trivial lower bound on the temperature
above which the thermal state is guaranteed to be separable.

A quantum system with Hamiltonian, $H$, in thermal equilibrium at
temperature, $T$, is described by the thermal state
\begin{equation}
  \rho_T = \exp(-\beta H)/Z\,,
\end{equation}
where $\beta = 1/ k_B T$ is the inverse temperature, $k_B$ is
Boltzmann's constant, and $Z = \tr[\exp(-\beta H)]$ is the partition
function. The energy of the thermal state, the \emph{thermal energy},
is given by
\begin{equation}
  U(T) = \tr[H \rho_{T}] = - \frac{1}{Z} \pderiv{Z}{\beta}\,. 
\end{equation}

\medskip\noindent
\textbf{Definition:} Given a system with an entanglement gap greater
than zero, $G_E > 0$, we define the \emph{entanglement-gap
  temperature}, $T_E$, to be the temperature at which the thermal
energy equals the minimum separable energy, $U(T_E) =
E_{\mathrm{sep}}$.\medskip

The thermal energy is a monotonically decreasing function of $\beta$
(i.e., it decreases as the temperature decreases).  By definition, all
states with energy less than $E_{\mathrm{sep}}$ are guaranteed to be
entangled, and thus the system is certainly entangled below the
entanglement-gap temperature.  That is, if we cool our system down
below the entanglement-gap temperature we know it must be in an
entangled state.  The thermal energy of the system, which depends only
on the temperature, becomes a witness to the entanglement of the
thermal state.

In order to compare Hamiltonians with different total energy ranges,
$E_{\rm tot}$, it is sensible to define a \emph{scaled temperature},
$t$ as
\begin{equation}
t = k_B T / E_{\rm tot}.
\end{equation}
The corresponding \emph{scaled entanglement-gap temperature} is $t_E =
k_B T_E / E_{\rm tot}$.

For the class of Hamiltonians identified in Theorem \ref{thm:
  maxentgap} with maximal entanglement gap, i.e., $H = I -
\qmproj{\Psi_{\rm me}}$, where $\ket{\Psi_{\rm me}}$ is a maximally
entangled state by the measure (\ref{eq:maxent}), it is
straightforward to calculate the entanglement-gap temperature.  The thermal energy is given by
\begin{equation}
  U(t) = \frac{(d_T -
  1)\exp(-\beta)}{1+(d_T-1) \exp(-\beta)}\,. 
\end{equation}
Setting $U(t_E) = E_{\mathrm{sep}} = M^{\rm max}$ gives
\begin{equation}
  t_E = \Bigl[\log_e \frac{(d_T-1)(1-M^{\rm max})}{M^{\rm
  max}}\Bigr]^{-1} \, .
\end{equation}

As an example, we consider the entanglement-gap temperature of a
bipartite system (each subsystem of dimension $d$), with Hamiltonian
$H = I - \qmproj{\phi_d}$ and scaled entanglement gap $g_E = 1 - 1/d$.
The scaled entanglement-gap temperature for this system is
\begin{equation}
  \label{eq: enttempmaxd} 
  t_E = \left[ \log_e (d+1) \right]^{-1}\,.  
\end{equation}
Note that the entanglement-gap temperature decreases with increasing
dimension despite the fact that the entanglement gap increases.  This
behavior is due to the fact that the number of eigenstates with
energy one increases quadratically with $d$, while the ground state
remains non-degenerate.

\subsection{Hamiltonians of Bipartite Systems Possessing Large Entanglement-Gap Temperature} 

It is natural to ask which Hamiltonians have the highest (scaled)
entanglement-gap temperature.  Somewhat counterintuitively it is not
the Hamiltonians with the largest entanglement gap.  In fact, there
are Hamiltonians with arbitrarily high entanglement-gap temperature.
To provide an example, we restrict our attention to the case where the
two subsystems of the bipartite system have the same dimension, $d_A =
d_B = d$.  The projectors onto the symmetric and antisymmetric
subspaces of $\mathcal{H}_A \otimes \mathcal{H}_B$ are $ \Pi_S = (I +
V_{(A,B)})/2$, and $\Pi_A = (I - V_{(A,B)})/2$, respectively, where
$V_{(A,B)}$ is the permutation operator on the two subsystems, defined
by $V_{(A,B)} \ket{\psi}_A \ket{\phi}_B = \ket{\phi}_A \ket{\psi}_B$
for all $\ket{\psi},\ket{\phi}$.  The antisymmetric subspace contains
only entangled states.  Thus if we define a Hamiltonian as the
projector onto the symmetric subspace
\begin{equation}
  \label{eq:SymProj}
  H = \Pi_S \,,
\end{equation}  
then all symmetric states have energy one, all antisymmetric states
have energy zero, and there is a finite entanglement gap.  We find the
gap by directly calculating the energy of a pure separable state,
$\ket{A} \ket{B}$,
\begin{equation} 
  \bra{A} \bra{B} H \ket{A} \ket{B} = (1 + |\braket{A}{B}|^2)/2 \,.
\end{equation}
From this expression it is clear that the minimum energy of $1/2$ is
achieved by any pure separable state such that $\braket{A}{B} = 0$.
The entanglement gap is $G_E = 1/2$, independent of $d$.

For the symmetric-projector Hamiltonian the thermal energy is given by
\begin{equation}
U(t) = \frac{d(d+1) \exp(-\beta)}{d(d-1)+d(d+1) \exp(-\beta)}
\end{equation}
Using $E_{\mathrm{sep}} = 1/2$ we find
\begin{eqnarray}
  \label{eq: symenttemp} 
  t_E &=& \left[ \log_e \left( \frac{d+1}{d-1} \right) \right]^{-1}, \\
  \nonumber &\simeq& d/2, \quad \mbox{for} \quad d \gg 1\,.  
\end{eqnarray}
Remarkably, for this Hamiltonian the scaled entanglement-gap
temperature increases without bound as the dimension of the subsystems
increases.

Thus, for Hamiltonians that only have eigenvalues zero or one, there
is a trade-off between ground-state degeneracy and the entanglement
gap in determining the entanglement-gap temperature.  Even though the
Hamiltonian with a non-degenerate maximally-entangled ground state has
a larger entanglement gap, the symmetric-projector Hamiltonian has a higher
entanglement-gap temperature due to its large ground-state degeneracy.

In Appendix~\ref{app: enttemp}, we investigate other Hamiltonians with
ground-state manifolds containing only entangled states, and present
evidence that no other bipartite Hamiltonian with a two-level energy
spectrum possesses an entanglement-gap temperature greater than the
Hamiltonian~(\ref{eq:SymProj}).  In Appendix~\ref{sec:
  maxenttempmegs}, we investigate the entanglement temperature of two
qubit systems, and provide evidence that the Heisenberg
antiferromagnetic Hamiltonian has the highest entanglement-gap
temperature.

We note that T\'{o}th \cite{TC} gives an example of a multiparty
Hamiltonian, the Heisenberg interaction between all pairs of $n$
spin-$1/2$ particles, whose entanglement-gap temperature increases
linearly with $n$, i.e. it is arbitrarily high for arbitrarily large
systems.  However, unlike our example, the total energy range also
increases linearly with $n$.  The scaled entanglement-gap temperature
of their Hamiltonian therefore approaches a constant as $n \rightarrow
\infty$.  By contrast, the entanglement-gap temperature of our example
is arbitrarily high despite the fact that the total energy range is
bounded.

\section{The Entanglement Gap of Quantum Many-Body Systems}
\label{sec: multipartite}

In this section, we investigate the entanglement gap for quantum
systems arranged on some graph or lattice that interact with some
local neighborhood.  Because we are only considering
finite-dimensional systems, the subsystems can always be thought of as
spins of some total angular momentum, so we use the terms
``subsystem'' and ``spin'' interchangeably.  For a particular type of
coupling -- bipartite lattices -- we provide an explicit method for calculating the entanglement gap, which applies to various spin systems often
considered in the condensed matter literature.  We also prove that, as
the coordination number grows, the entanglement gap per interaction
must decrease to zero.  This result makes use of the fact that as the
number of equivalent spins connected to a given spin in the lattice
increases there does not exist a global state of the system for which
each interacting pair is strongly entangled.

\subsection{The Entanglement Gap of 2-Local Hamiltonians}

\label{sec: entgap2loc}

We now consider multipartite systems with only two-body interactions.
The Hamiltonian for such a system can be defined by a set of coupling
Hamiltonians $H_{ij}$ that act as the identity on all the spins other
than $i$ and $j$, and a graph or lattice where the vertices represent
spins and edges represent an interaction between the spins on the two
sites.  We refer to each two-body interaction, or edge on the graph, as a
\emph{bond}.  The Hamiltonian for the entire system is
\begin{equation}
  \label{eq: graphhamil} 
  H = \sum_{<i,j>} H_{ij}\,,
\end{equation}
where $\sum_{<i,j>}$ indicates a sum over vertices connected by an
edge, i.e., a sum over bonds.  We refer to such a Hamiltonian as
\textit{2-local}.  It follows that that the energy of a 2-local
Hamiltonian depends only on the reduced density matrices of each
interacting pair; see e.g.~\cite{lidar}.  Thus, the energy for any
global state $\rho$ is
\begin{equation}
  \label{eq:reddens}
  E=\tr [H\rho] =\sum_{<i,j>} \tr [H_{ij}\rho_{ij}] \,,
\end{equation}
where $\rho_{ij}$ is the reduced state of the interacting pair of
spins ${<}i,j{>}$.  In the following we consider only systems where
each of the coupling Hamiltonians $H_{ij}$ is equal.

We note that the reduced states $\rho_{ij}$ are not completely
arbitrary if they are to be consistent with a global state $\rho$ for
the whole system.  In particular, the ground state of the graph or
lattice cannot simply be constructed from the reduced states that
minimize the energy of each bond ($\tr [H_{ij}\rho_{ij}]$), because
these reduced states may not be consistent with a global state.  As we
demonstrate below, this situation can occur when there is a non-zero
entanglement gap for the coupling Hamiltonian $H_{ij}$.  Motivated by
the results of Sec.~\ref{sec: hamentwit} as well as its importance in
condensed matter physics, we use the Heisenberg antiferromagnet as our
standard example.

A \emph{bipartite} graph or lattice is one for which the vertices can
be divided into two sets, $A$ and $B$, such that the edges only
connect vertices from $A$ with vertices from $B$.  Examples of
bipartite lattices include the square lattice, see Fig.~\ref{fig:
  lattice}, and the hexagonal lattice on the plane.  An even number of
vertices arranged in a ring is also bipartite.  

For bipartite graphs or lattices, we now demonstrate how to construct a separable state with the lowest possible energy.  First, consider a minimum-energy
separable state $\ket{A} \ket{B}$ for a pair of spins under the
interaction Hamiltonian, and construct a global state, $\otimes_{i_A
  \in A} \ket{A}_{i_A} \otimes_{i_B \in B} \ket{B}_{i_B}$, such that
all the spins on the subset $A$ are in the state $\ket{A}$ and
likewise for $B$.  By Eq.~(\ref{eq:reddens}) the energy per bond of
this state is the same as the energy $E_{\mathrm{sep}}$ of the state
$\ket{A} \ket{B}$, and it provides an upper bound on the minimum
separable energy per bond of the full Hamiltonian (\ref{eq:
  graphhamil}).  To see that there is no global separable state with
lower energy, suppose that such a state exists.  Then all of the
nearest neighbor reduced density matrices $\rho_{ij}$ must be
separable and by Eq.\ (\ref{eq:reddens}) at least one of them must
have a lower energy under the interaction Hamiltonian than $\ket{A}
\ket{B}$: a contradiction. Therefore, the state $\otimes_{i_A \in A}
\ket{A}_{i_A} \otimes_{i_B \in B} \ket{B}_{i_B}$ is indeed a
minimum-energy separable state of the entire system.

\begin{figure}[htb]
  \includegraphics[width=80mm]{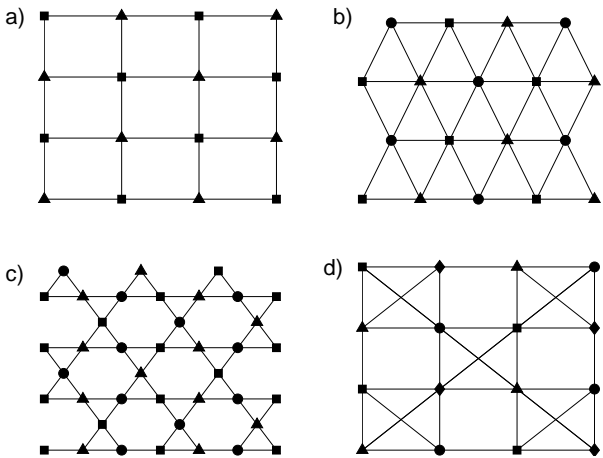}
  \caption{Examples of $n$-partite lattices.  (a) square lattice ($n=2$),
    (b) triangular lattice ($n=3$), (c) kagom\'{e} lattice ($n=3$), (d)
    checkerboard lattice ($n=4$).  The $n$ different markers indicate
    the $n$ subsets that the vertices of the $n$-partite lattice may
    be divided into so that there are only interactions between
    distinct subsets.}
\label{fig: lattice}
\end{figure}  

Given this result, we can determine the lowest possible energy of a
separable state for the Hamiltonian (\ref{eq: graphhamil}) on any
bipartite graph or lattice simply by solving the problem for a single
pair of spins. Finding the entanglement gap for such systems reduces
to finding the entanglement gap of the two-body interaction and the
ground-state energy $E_0$ of the overall system. This important fact
was noted by T\'oth \cite{TC} who provided a slightly different
argument.

Consider a bipartite graph with Hamiltonian $H$.  Then our argument
proves that the operator $H-NE_{\mathrm{sep}}$ is an entanglement
witness, where $N$ is the number of bonds.  Note that we can express
this entanglement witness as a sum over bonds,
\begin{equation}
  H-NE_{\mathrm{sep}}=\sum_{<i,j>} (H_{ij}-E_{\mathrm{sep}})\,, 
\end{equation}
where each term in the sum is a bipartite entanglement witness.  As a
result the expectation value depends only on the bipartite reduced
density matrices of nearest neighbors and can only be negative if
these reduced density matrices are entangled.  This result can be
extended to apply to lattices (with $N \to \infty$) as well.  So,
while the ground-state energy is certainly a global quantity, this
construction is only sufficient to detect the existence of bipartite
entanglement between interacting pairs in a bond.

In a similar way we can calculate the entanglement gap for $n-$partite
graphs or lattices that are formed by groups of $n$ spins each having
an ``all-to-all'' interaction graph.  We first construct a
minimum-energy separable state of a single group of $n$ spins with an
all-to-all interaction graph.  This state extends to minimum-energy
separable state of the entire lattice as depicted in Fig.~\ref{fig:
  lattice}.  We then compare the minimum-separable energy per bound to
the ground state energy per bond to find the entanglement gap per
bond.  For example, on the tripartite triangular and kagom{\'{e}}
lattices (see Fig.~\ref{fig: lattice}), it is possible to calculate
the entanglement gap from the ground-state energy and the minimum
separable energy of three spins having an all-to-all interaction graph
(a single triangle).  We describe the nature of the entanglement that
can be witnessed in these systems in Sec.~\ref{sec: frustration}.

\subsection{Entanglement Gap and coordination Number}

The \emph{coordination number} of a lattice is the number of edges
incident on each lattice site, i.e., the number of other systems that
each spin interacts with via the coupling Hamiltonian
$H_{ij}$.  As we are now considering lattices, our assumption that all
interactions $H_{ij}$ are equal implies translational symmetry.  We
now investigate how the entanglement gap varies with the coordination
number of the lattice.  The basic idea stems from the fact that, as a
result of the translational symmetry, the ground state of our 2-local
Hamiltonians have equal reduced density matrices for interacting
pairs~\footnote{If there is spontaneous symmetry breaking a mixture of
  the symmetry-broken ground states will be translationally
  invariant.}.  As the coordination number increases, this equality
requires that every spin shares the same reduced density matrix with
an increasing number of other spins.  The results of
\cite{Doherty2004a,raggio1989a,fannes1988a} then preclude the reduced
density matrices from being strongly entangled.  Building on these
results, we prove a theorem stating that, as the coordination number
of the lattice grows, the entanglement gap decreases to zero. We then
investigate this behaviour in the specific case of the Heisenberg
antiferromagnet.

In order to prove results about the maximum possible entanglement gap
in Sec.~\ref{subsec: bientgap} it was natural to use the scaled
entanglement gap.  However in what follows it is more convenient to
use the \emph{entanglement gap per bond}, $G_E/N$, where $N$ is the
total number of bonds.  (Note that this entanglement gap per bond is
well defined even for lattices with $N \to \infty$.)  These two
methods of scaling are roughly equivalent because the total energy
range of the system tends to scale linearly with the number of sites.

We begin by considering a restricted set of graphs which we will use
to prove results that bound the entanglement gap on any bipartite
lattice.  We define a \emph{star} graph as a bipartite graph where
there is only a single vertex, the \emph{centre}, in one subset, $A =
\{ A_0\}$, and $k$ vertices, the \emph{points}, in the other subset,
$B = \{B_i , i = 0, \ldots, k-1\}$, and where edges connect each point
and the centre.  

A \emph{strictly positive} entanglement witness, $Z_{EW}$, is a
Hermitian operator whose average is strictly positive on separable
states, $\tr [Z_{EW} \rho_{\mathrm{sep}}] > 0, \quad \forall\ 
\rho_{\mathrm{sep}} \in \mathcal{S},$ but which has at least one
negative eigenvalue.

Before stating and proving our main theorem we present the following
lemma:

\begin{lem}
\label{lem: symposentwit}
Let $Z_{EW}$ be a stictly positive entanglement witness acting on
$\mathcal{H}_{A_0} \otimes \mathcal{H}_{B_0}$.  Then there exists a
positive integer $k$ such that
\begin{equation}
  \sum_{i=0}^{k-1} V_{(B_0,B_i)} (Z_{EW} \otimes_{i=1}^{k-1} I_{B_i})
  V_{(B_0,B_i)}^\dagger \geq 0\,,
\end{equation} 
where $V_{(B_0,B_i)}$ is the self-adjoint unitary operator that swaps
the Hilbert spaces $\mathcal{H}_{B_0}$ and $\mathcal{H}_{B_i}$.
\end{lem}

This Lemma is a straightforward modification of Theorem 2 of
\cite{Doherty2004a} and the proof follows similarly~\footnote{The key
  difference between this proof and the one found
  in~\cite{Doherty2004a} is that here, mixing with the operators
  $V_{(B_0,B_i)}$ forces the resulting entanglement witness to be
  block-diagonal in the irreducible representations of the symmetric
  group, rather than projecting into the fully symmetric
  representation of this group as in~\cite{Doherty2004a}.}.

Using the general mapping between entanglement witnesses and
Hamiltonians with non-zero entanglement gap discussed in Sec.\ 
\ref{sec: hamentwit}, this result on strictly positive entanglement
witnesses bounds the entanglement gap for Hamiltonians on star graphs.

\begin{thm}
\label{thm: starentgap}
For any coupling Hamiltonian $H_{A_0 B_0}$ and any $\epsilon > 0 $ there
exists a positive integer, $k$, such that the entanglement gap per
interaction for the Hamiltonian (\ref{eq: graphhamil}) on a star graph
with $k$ points is less than $\epsilon$.
\end{thm}

\noindent
\textbf{Proof:} The non-trivial case occurs when $H_{A_0 B_0}$ has
non-zero entanglement gap.  Note that the total Hamiltonian on the
star graph may be written as
\begin{equation}
  H^{\mathrm{star}} = \sum_{i=0}^{k-1} V_{(B_0,B_i)} (H_{A_0 B_0}
  \otimes_{i=1}^{k-1} I_{B_i}) V_{(B_0,B_i)}^\dagger \,.
\end{equation} 
We define a strictly positive entanglement witness on
$\mathcal{H}_{A_0} \otimes \mathcal{H}_{B_0}$ as
\begin{equation}
  Z_{EW} = H_{A_0 B_0} - E_{\mathrm{sep}}I + \epsilon I \,, 
\end{equation} 
where $E_{\mathrm{sep}}$ is the energy of the minimum-energy separable
state of $H_{A_0 B_0}$ and, by adding $\epsilon > 0$, $Z_{EW}$ is
guaranteed to be a strictly positive entanglement witness.  From Lemma
\ref{lem: symposentwit}, there exists a $k$ such that
\begin{equation*}
  \sum_{i=0}^{k-1} V_{(B_0,B_i)} \left( ( H_{A_0 B_0} - E_{\mathrm{sep}} +
  \epsilon) \otimes_{i=1}^{k-1} I_{B_i} \right) V_{(B_0,B_i)}^\dagger
  \geq 0\,,
\end{equation*} 
and so $H^{\mathrm{star}} \geq k (E_{\mathrm{sep}} - \epsilon)$.
Because the energy of the minimum-energy separable state of
$H^{\mathrm{star}}$ is $k E_{\mathrm{sep}}$, this implies that the
entanglement gap of the total Hamiltonian satisfies
$G_E^{\mathrm{star}} \leq k \epsilon$.  Thus, given any $\epsilon > 0$
there exists a $k$ such that $G_E^{\mathrm{star}} / k \leq \epsilon$,
as claimed.  \hfill$\Box$\medskip

As an illustration of this theorem we consider the spin-$1/2$
Heisenberg antiferromagnetic Hamiltonian on a star graph.  Recall that
the coupling Hamiltonian is $H_{ij} = \vec{\sigma_i} \cdot \vec{\sigma_j}$;
the entanglement gap of this coupling Hamiltonian was investigated in
Sec.~\ref{subsec: bientgap}.  The ground state is the singlet, with
energy $-3$, and the three triplet states all have energy $+1$.  The
minimum-energy separable states are of the form $\ket{A} \ket{B}$ such
that $\braket{A}{B} = 0$, with energy $-1$.

Using the permutation symmetry amongst the points of the Hamiltonian
on the star graph it is possible to calculate its ground-state energy
exactly~\cite{weiss1948}, $E_0=-(k+2)$.  The coordination number of the
center of the graph is the number of points, $k$.  The energy of any
minimum-energy separable state is $E_{\mathrm{sep}}=-k$.  In Table
\ref{tab: starentgap} we present the entanglement gap per bond and the
scaled entanglement gap for comparison with other lattices below.
\begin{table}[htb]
\begin{tabular}{|c|c|c|c|c|c|} \hline
Coord. & $E_0$ per & $E_{\mathrm{sep}}$ per & Ent. Gap & Scaled  \\
  No. $k$ &  bond & bond & per bond& Ent. Gap \\ \hline \hline
  1& -3 &-1& 2& 0.5  \\
  2 & -2 &-1 & 1 & 0.333\\
 3 & -1.667 &-1& 0.667 & 0.25 \\
 4& -1.5 &-1 & 0.5 & 0.2\\
 5& -1.4 &-1 & 0.4 & 0.167\\
 6&  -1.333 &-1 & 0.333 & 0.143\\ \hline
\end{tabular}
\caption{Properties of star graphs with the Heisenberg antiferromagnetic
  Hamiltonian as a function of coordination number $k$.  The ground-state
  energy, minimum separable energy and entanglement gap are
  all \emph{per bond}, i.e., energies divided by $k$. The scaled
  entanglement gap is the entanglement gap divided by the total energy
  range of the system.} 
\label{tab: starentgap}
\end{table}

Although the Heisenberg antiferromagnet has the largest entanglement
gap for two qubits, we have not proved that it has the largest
entanglement gap per bond on a star graph.  However we have calculated
the entanglement gap per bond for numerous common spin models such as
the $XXZ$ model and $XY$ model, all of which have a smaller
entanglement gap per bond.  Therefore we conjecture that the
Heisenberg antiferromagnet has the largest entanglement gap per bond
on a star graph.  If this were true it would provide an upper bound
of $O(1/k)$ on the approach to zero of the entanglement gap per bond
implied by Theorem \ref{thm: starentgap}.

In order to determine the entanglement gap on a bipartite lattice, we
require the ground-state energy of the lattice as well as the lowest
energy achievable by a separable state of a single pair of spins, as
noted above.  The ground-state energy of a star graph can be used to
bound the ground-state energy of a bipartite lattice with the same
coordination number as follows.

\begin{lem}
\label{lem: bistar}
The ground-state energy per bond of any coupling Hamiltonian on a
bipartite lattice with coordination number $k$ is greater than or
equal to the ground-state energy per bond of the same Hamiltonian on a
star graph with $k$ points.
\end{lem}

\noindent
\textbf{Proof:} The essential idea is to divide the expression for
ground-state energy on the lattice into a sum over star graphs with
$k$ points where $k$ is the coordination number of the lattice.  Let
$\rho_0$ denote the translationally-invariant ground state of the
entire lattice.  Consider the star graph consisting of a particular
lattice site (the center) and those sites connected to it by a 
coupling (the points).  The reduced state on the star graph is
obtained by tracing out all sites not in the star
\begin{equation}
  \rho_{\mathrm{star}} = \tr_{i \not\in \{ \mathrm{star}
  \}}[\rho_0]\,. 
\end{equation}     
This state is independent of the lattice site chosen as the center
(due to the translational invariance of $\rho_0$), and the energy per
bond of this reduced state is the same as the ground-state energy per
bond of the lattice.  The ground-state energy is then
\begin{equation}
  \label{eq:grdstar}
  E_0=\tr [H\rho_0 ] =\sum_{i}
  \tr [H_{\mathrm{star}} \rho_{\mathrm{star}}]/k \,,
\end{equation}
Furthermore, the energy of $\rho_{\mathrm{star}}$ can only be greater
than the energy of a ground state $\ket{E_0}_{\mathrm{star}}$ of the
star Hamiltonian,
\begin{equation}
  \tr [H_{\mathrm{star}} \rho_{\mathrm{star}}] \geq \tr[
  H_{\mathrm{star}} \ket{E_0}_{\mathrm{star}} \bra{E_0}] \,.
\end{equation}   
It follows that the ground-state energy per bond of the bipartite
lattice is greater than or equal to the ground-state energy per bond
of the star graph. \hfill$\Box$\medskip

We note that a similar argument is used in \cite{A51} to bound the
ground-state energy of the Heisenberg antiferromagnet.

Using this bound for the ground-state energy, it is straightforward to
bound the entanglement gap on bipartite lattices, which is the main
result of this section.

\begin{thm}
\label{thm: bicoordentgap}
Given any $\epsilon > 0 $ there exists a positive integer, $k$, such
that the entanglement gap per bond for an arbitrary coupling Hamiltonian
on any bipartite lattice with coordination number $k$ is less than
$\epsilon$.
\end{thm}

\noindent
\textbf{Proof:} Because the bipartite lattice and star graph are both
bipartite, they have the same minimum separable energy per bond. The
result now follows from Theorem \ref{thm: starentgap} and Lemma
\ref{lem: bistar}. \hfill$\Box$\medskip

\begin{table}[htb]
\begin{tabular}{|c|c|c|c|c|c|} \hline
Lattice & Coord. & $E_0$ per & $E_{\mathrm{sep}}$ per & Ent. Gap & Scaled  \\
 & No. &  bond & bond & per bond& Ent. Gap \\ \hline \hline
single bond & 1& -3& -1 & 2 & 0.5 \\ \hline
$1D$ chain & 2 & -1.772 & -1& 0.772 & 0.279 \\
hexagonal& 3 & -1.452 & -1 & 0.452 & 0.184 \\
square & 4& -1.338 & -1& 0.338 & 0.145 \\
cubic & 6& -1.194$^*$ & -1 & 0.194 & 0.088 \\ \hline
single triangle & 2 & -1 & -0.5 & 0.5 & 0.25 \\ \hline
kagom\'e & 4 & -0.874 & -0.5 &0.374 & 0.200\\ 
triangular & 6 & -0.726 & -0.5 & 0.226 & 0.131 \\ \hline
\begin{minipage}{0.6in}{single}\\{tetrahedron} 
\end{minipage}& \raisebox{0pt}[13pt][7pt]{3} &
 \raisebox{0pt}[13pt][7pt]{-1} & \raisebox{0pt}[13pt][7pt]{-0.333} &
 \raisebox{0pt}[13pt][7pt]{0.667}  & \raisebox{0pt}[13pt][7pt]{0.333}   
\\ \hline
checkerboard & 6& -0.67$^\dagger$& -0.333 & 0.34 & 0.20 \\ \hline
\end{tabular}
\caption{Entanglement gap for the Heisenberg antiferromagnet for
 various bipartite and frustrated lattices with different coordination
 numbers.  Ground-state energies taken from \cite{LM}, except $^*$ from linear
 spin-wave theory \cite{A52} and $^\dagger$ from \cite{fouet2003a}. } 
\label{tab: latticeentgap}
\end{table}

To illustrate this theorem, we calculate the entanglement gap per bond
of the spin-$1/2$ Heisenberg antiferromagnet on simple bipartite
lattices with varying coordination number.  In Table \ref{tab:
  latticeentgap} we present the ground-state energy, taken from the
literature, and thus the entanglement gap per bond and scaled
entanglement gap for a Heisenberg antiferromagnet on a $1D$ chain,
honeycomb, square and cubic lattice (all bipartite), as well as some
non-bipartite lattices to be discussed in Sec.~\ref{sec:
  frustration}.  It can be seen that the entanglement gap per bond
does decrease with increasing coordination number for the bipartite
case, and is always less than that of the corresponding star graph in
Table \ref{tab: starentgap}, as proved by Lemma \ref{lem: bistar}.
The entanglement gap per bond appears to decrease with coordination
number on tripartite lattices as well, thus providing evidence that
this behavior is not confined to bipartite lattices.

\section{Discussion}
\label{sec:Discussion}

In this section, we discuss some of the implications of our results
and explore the connections with other results from the condensed
matter literature.  We also discuss frustrated lattices and quantum
phase transitions and their effect on the entanglement gap.

The energy gap between the ground-state energy and the lowest energy
achieved by a separable state has been discussed in the quantum
magnetism literature using a slightly different terminology \cite{LM}.
There, separable states are associated with ``classical
configurations,'' arrangements of classical spin vectors minimizing
the energy of the appropriate classical Heisenberg spin model. The
reduction in ground-state energy below this point is typically
ascribed to ``quantum fluctuations.'' As a result, Table~\ref{tab:
  latticeentgap} is essentially drawn directly from the review by
Lhuillier and Misguich \cite{LM}.  Our results show that, in this
context at least, the term ``quantum fluctuations'' as discussed in
the condensed matter literature can be identified precisely with
entanglement as discussed in the quantum information literature, and
the associated reduction of ground-state energy in antiferromagnets
can be directly related to the theory of mixed state
entanglement~\cite{terhal2002a,lewenstein2000b,bruss2002a}.

Note that the entanglement gap is well over a quarter of the total
energy range for Heisenberg antiferromagnet on a line and is a
very significant fraction of the total energy range for the majority
of the lattices considered. This large entanglement gap reinforces the
argument made by Brukner and Vedral~\cite{BV} that the entanglement
witnesses resulting from the energy of appropriate spin models can
have macroscopic expectation values.

\subsection{Bipartite Lattices and Mean-Field Theory}

\label{sec: mft}

\emph{Mean-field theory} is a term used to describe a variety of
techniques in condensed matter physics for finding an approximation to
the ground state of a quantum many-body system.  Typically such
techniques correspond to searching for a separable state that
approximates the ground state.  It is a well-known observation that
mean-field theory is more accurate in higher dimensions and, because
coordination number typically increases with the dimension, for higher
coordination number. So for example, dynamical mean field theory for
fermion systems is known to be exact in infinite dimensions~\cite{georges1996a}.

In the context of our present work, we expect mean-field theory to
work well when the entanglement gap is small, because there exists a
separable state that has energy close to the ground-state energy and
thus a variational approach involving separable states might be
expected to be accurate.  Theorem \ref{thm: bicoordentgap}
demonstrates in a precise way that the entanglement gap decreases to
an arbitrarily small value with increasing coordination number on
bipartite lattices, independent of the particular coupling
Hamiltonian. This result is therefore suggestive of a quantitative
connection between entanglement and the improvement of mean-field
theory with dimension.

The work of Raggio and Werner~\cite{raggio1989a} aimed to develop a
rigorous mean field theory for Hamiltonian models on star graphs with
a large number of points.  Our results are ultimately based on a
characterization of bipartite separable states proven there and in
Ref.~\cite{fannes1988a}, subsequently used in Ref.~\cite{Doherty2004a}
to prove a result closely related to our Lemma~\ref{lem:
  symposentwit}.  The proofs in Ref.~\cite{raggio1989a} are
technically very difficult, because they apply not only to finite
dimensional spin systems but to any quantum system defined on a
separable Hilbert space.  These results may provide a more direct
route to our Theorem~\ref{thm: starentgap} for star-shaped graphs,
which could then be used to prove the result for bipartite lattices in
more generality; however, we have preferred to give a simple
derivation valid for finite-dimensional spin systems.

\subsection{Frustrated Lattices and Multipartite Entanglement}

\label{sec: frustration}

Lattices that are not bipartite lead to spin systems that are often
referred to as \emph{frustrated} in condensed matter physics
\cite{moessner2001a}.  This terminology arises from the fact that the
minimum-energy separable state for two neighboring sites on such a
lattice is not equal to the minimum-energy separable for the two sites
coupled alone.  As a result the energy per bond on such a lattice is
higher than the energy of a single pair for the same
interaction~\footnote{Dawson and Nielsen~\cite{dawson2004a} derive a
  bound on the ground-state entanglement based on the frustration of
  the \emph{quantum} Hamiltonian, not the frustration of its classical
  counterpart.}. The physics of frustrated quantum and classical spin
systems have been a subject of intensive research in recent years; we
refer the reader to~\cite{moessner2001a} for a review. In the
following, we briefly consider the effect of frustration on the
entanglement gap.

Further motivation for studying lattices that are not bipartite comes
from considering the nature of the entanglement detected by the
Hamiltonian.  On bipartite lattices, entanglement is only detected
when the reduced density matrices associated with each bond are
entangled.  So, for example, states which are multipartite entangled
but contain no bipartite entanglement, such as the three-party GHZ
state, will never have lower energy than the minimum separable energy
on a bipartite graph for any interaction Hamiltonian.  On
non-bipartite lattices it is sometimes possible for a coupling
Hamiltonian to witness the entanglement of such states.

A simple example of a non-bipartite lattice is the regular triangular
lattice, which is tripartite but not bipartite.  We consider two other
non-bipartite lattices in two dimensions: the kagom\'e lattice, which
is made up of corner-sharing triangles, and the checkerboard lattice.
These lattices are depicted in Fig.~\ref{fig: lattice}.

Again we consider the Heisenberg interaction.  In order to find
the lowest separable energy for the triangular and kagom\'e lattices,
we first find a minimum-energy separable state for a single triangle, as described in Sec.~\ref{sec: entgap2loc}.  The total
Hamiltonian for a single triangle is
\begin{equation}
  H = \vec{\sigma}_1 \cdot \vec{\sigma}_2+\vec{\sigma}_2 \cdot
  \vec{\sigma}_3+\vec{\sigma}_3 \cdot \vec{\sigma}_1 \,.
\end{equation}
Its spectrum and minimum-energy separable states may be found by
standard symmetry methods (for example~\cite{EW01}).  The ground state
is four-fold degenerate with energy $E_0 = -3$.  For the Heisenberg
antiferromagnet of $n$ spins with an all-to-all coupling, it is
straightforward to show that a minimum-energy separable state is given
by any configuration of spins where the total spin vector is
zero~\cite{moessner2001a}.  For the triangle, a minimum-energy
separable state is
\begin{equation}
  \ket{\uparrow}_1 \otimes (\ket{\uparrow}_2 + \sqrt{3}
  \ket{\downarrow}_2)/2\otimes(\ket{\uparrow}_3 - \sqrt{3}
  \ket{\downarrow}_3)/2 \,,
\end{equation}
which corresponds to a classical configuration of spins at an angle of
$2 \pi/3$ from each other in the plane having a total spin zero (the
``Mercedes star'' configuration in~\cite{EW01}).  This state has
energy $E_{\mathrm{sep.}} = -3/2$.  The maximum energy manifold is
spanned by the states with all spins parallel, and has
energy $E_{\mathrm{max}} = 3$.

From these results we can calculate the entanglement gap per bond and
the scaled entanglement gap for the Heisenberg interaction on the
triangle, shown in Table~\ref{tab: latticeentgap}.  Also shown are the
entanglement gaps for the kagom\'e and triangular lattices, calculated
from $E_{\mathrm{sep}}$ for the triangle, and the ground-state energy
of the entire lattice, taken from~\cite{LM}.  Note that, as for
bipartite lattices, the entanglement gap per bond appears to decrease
with coordination number.  We have also considered the checkerboard
lattice (see Fig.~\ref{fig: lattice}) which is made up of
corner-sharing tetrahedra and has a coordination number of six. We
obtain the ground-state energy of this model from
Ref.~\cite{fouet2003a}, where it is estimated from exact
diagonalization of small samples.

The reduced density matrices associated with bonds of the lattice in
the ground state are not entangled for these frustrated systems.  Note that the symmetries of the Heisenberg model guarantee that these bipartite reduced
density matrices are so-called Werner states~\cite{werner1989a},
invariant under any local unitary rotation of the form $U\otimes U$.
These states are entirely characterized by the fraction of the
population that is in the singlet state, and when this fraction is
less than a half the state is separable~\cite{werner1989a}.  Therefore the reduced density matrix associated with each bond is separable
whenever the ground-state energy per bond is above the minimum
separable energy of the Heisenberg Hamiltonian for a single pair of
spins. With the ground-state energy per bond from Table~\ref{tab:
  latticeentgap}, it is clear that there is no bipartite entanglement
of nearest neighbor spins for the Heisenberg model on the triangular,
kagom\'e or checkerboard lattices (because $E_0$ per bond is greater than -1). The entanglement gap for these
systems is associated with the entanglement of the reduced states of
the triangles or tetrahedrons that make up the lattice.  Thus, the
Hamiltonian serves as a witness for multipartite entanglement in these
systems.

It appears that as the frustration of the classical spin model
increases, so does the entanglement gap. For a coordination number of
six the entanglement gap as a fraction of the overall energy range of
the Hamiltonian increases from $0.088$ on the bipartite cubic lattice
to $0.131$ on the tripartite triangular lattice and finally to around
$0.2$ on the checkerboard lattice.  It would be interesting to
understand this behavior in more detail.  It is a feature of
frustrated classical spin models that they have a large number of
configurations achieving the lowest possible energy, which may be a
contributing factor to this observed larger entanglement gap.

\subsection{The Entanglement Gap in a Simple Quantum Phase Transition}

\label{sec: QPT}

The role of entanglement in \emph{quantum phase transitions} \cite{Sac} is currently of considerable interest \cite{ON02,OAF+02,lidar,VLR+03}.  Perhaps the simplest model to exhibit a quantum phase transition, used in many of these studies, is the $1D$ infinite-lattice transverse field $XY$ model with Hamiltonian
\begin{equation}
  \label{eq: XYhamil1D} 
  H = \sum_{j=0}^{N-1} \left( \frac{1 + \gamma}{2}
  \sigma^x_j \sigma^x_{j+1} + \frac{1 - \gamma}{2} \sigma^y_j
  \sigma^y_{j+1}  + \lambda \sigma^z_j  \right) \,,
\end{equation}
where $\gamma$ is the anisotropy in the $x-y$ plane, and $\lambda$ is
an external magnetic field, $N$ is the total number of lattice sites,
and cyclic boundary conditions are imposed so that a subscript $N$ is
identified with $0$.  For $\gamma = 1$ the transverse field Ising
model is recovered.

It is of interest to see how this phase transition affects the
entanglement gap.  Here we calculate the entanglement gap of the $1D$
$XY$ model as a function of $(\gamma,\lambda)$ in the thermodynamic
($N \rightarrow \infty$) limit.  Because a $1D$ lattice is bipartite
(for $N$ even), given knowledge of the ground-state energy it is
sufficient to calculate the entanglement gap for the coupling Hamiltonian
in order to calculate the entanglement gap of the entire system, as
described in Sec.~\ref{sec: multipartite}. In this case the coupling Hamiltonian may be chosen to be
\begin{equation}
  \label{eq: XYhamil} 
  H^{XY}_{ij} = \frac{1 + \gamma}{2} \sigma^x_i \sigma^x_j + \frac{1 -
  \gamma}{2} \sigma^y_i \sigma^y_j  + \frac{\lambda}{2} (\sigma^z_i +
  \sigma^z_j) \,, 
\end{equation}
where the factor of $1/2$ in front of the magnetic field accounts for
the fact that each site is involved in two interactions.  In
Appendix~\ref{app: TFXY} we calculate the minimum-separable energy for
this coupling Hamiltonian, Eq.~(\ref{eq: EsepXY}).

The $XY$ model on a $1D$ chain, Eq.~(\ref{eq: XYhamil1D}), is
well-known to be exactly solvable via the Jordan-Wigner
transformation; see e.g.~\cite{Sac}.  We obtain the ground-state
energy from this method.

\begin{figure}[htb]
  \includegraphics[width=80mm]{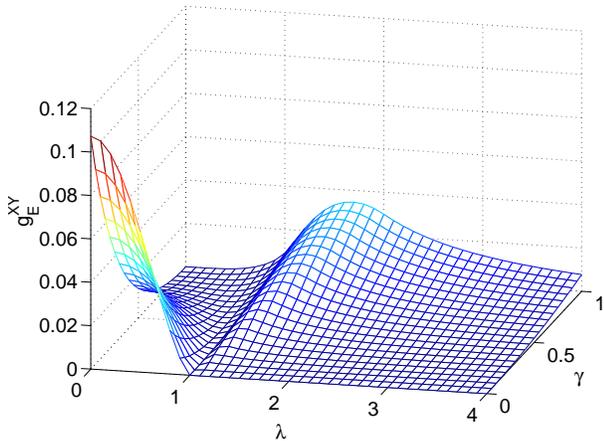}
\caption{Entanglement gap as a function of anisotropy $\gamma$ and
  transverse field $\lambda$ for $XY$ Hamiltonian on a $1D$ lattice in
  the thermodynamic limit.} 
\label{fig: XYgapNlarge}
\end{figure}

In Fig.~\ref{fig: XYgapNlarge} we plot the scaled entanglement gap as
a function of $(\gamma,\lambda)$ in the thermodynamic limit.  The
quantum phase transition in this model occurs at $\lambda = 1$ for
$\gamma \neq 0$.  Previous studies have indicated that the ground
state becomes highly entangled at this point, and this behavior is
manifest in a sudden rise in the entanglement gap about this point.
Intuitively one might expect that the more entangled the ground state,
the larger the entanglement gap.  While qualitatively true, this connection cannot be exact
because the entanglement gap is a property of the whole Hamiltonian;
it can depend on all energy eigenstates and their energies and is not
just a property of the ground state.

Nevertheless, given the discussion in Sec.~\ref{subsec: bientgap}
it is reasonable to consider a connection between the ground-state
entanglement as measured by $M(|\Psi\rangle)$ of Eq.~(\ref{eq:maxent})
and the entanglement gap. The entanglement of the ground state under
this measure has been investigated in the $XY$ model in recent work by
Wei \emph{et al.}~\cite{WDM+}.  This measure depends only on the
maximum overlap of the entangled state $\ket{\Psi}$ with a separable
state.  One might expect that the minimum-energy separable state is
one which has the maximum overlap with the ground state.  However, in order to achieve maximum overlap with the
ground state it may also be necessary for a separable state to have
large overlap with high-energy eigenstates.  Therefore a
separable state may achieve lower energy by having less overlap with
the ground state but considerably more overlap with low-lying excited
states.  

In Ref.~\cite{WDM+} the derivative of the global entanglement with
respect to the external field was found to contain a singularity at
the critical point consistent with the universality class of the
model.  Although we see a qualitatively similar peak near the critical
point, there is no singularity in the derivative of the entanglement
gap.  Again, such a singularity may not have been expected because the
entanglement gap is not simply a property of the ground state.

\subsection{Summary}

We have studied entanglement in quantum many-body systems from the
point of view of the Hamiltonian as an entanglement witness.  We
introduced two related concepts useful in studying the role of
entanglement in the ground and thermal states of multipartite quantum
systems.  The first is the entanglement gap, which is the difference
in energy between the ground-state energy and the minimum energy that
any separable state can attain.  If the energy of the system lies
within the entanglement gap range, the state of the system is
guaranteed to be entangled.  The second concept is the
entanglement-gap temperature: the temperature at which the
energy of the thermal state is equal to the minimum separable energy,
and below which the thermal state must be entangled.  The
entanglement-gap temperature provides a threshold for deducing the
thermal state of the system to be entangled, based on its energy.

For multipartite, finite-dimensional quantum systems we proved that
Hamiltonians possessing a non-degenerate maximally entangled ground
state (according to a global measure of entanglement) and all other
energy eigenstates degenerate maximize the entanglement gap.  The
related question of which Hamiltonians have the highest entanglement
gap temperature is more challenging; substantial evidence is given
that the Heisenberg antiferromagnetic Hamiltonian has the largest
entanglement temperature for two qubits.

On bipartite lattices, i.e., those lattices for which there are only
interactions between two disjoint subsets of the vertices, we proved
that the entanglement gap decreases to zero as the coordination number
increases.  This result suggests a quantitative reason why
approximation schemes based on separable states, such as various forms
of mean-field theory, appear to give more reliable results at higher
coordination number.

On frustrated lattices, i.e., those that are not bipartite, we noted
that the Hamiltonian can act as an entanglement witness for
multipartite entanglement, even when there is no bipartite
entanglement present.  Finally, we calculated the entanglement gap
near a simple quantum phase transition, and showed that although it
does not follow any universal scaling law, it does increase near the
quantum phase transition, as may have been expected from previous
studies in which the ground state was found to become highly entangled
at that point.

\begin{acknowledgments}
  We thank Jennifer Dodd, Aram Harrow, Michael Nielsen and Ben Powell
  for helpful discussions.
\end{acknowledgments}

\appendix

\section{Semidefinite programs for the entanglement gap}
\label{app: semidefentgap}

We will now describe efficient numerical procedures for evaluating the
entanglement gap of a given Hamiltonian using semidefinite programs.

Semidefinite programs are a type of convex optimization problem
\cite{VB96,VBbook}, which are appealing because they have efficient
numerical implementations.  With the view of Hamiltonians as
entanglement witnesses, and following methods described in
\cite{DPS02,Doherty2004a,doherty2004b} it is possible to express the
problem of finding the minimum separable energy as a sequence of
semidefinite programs, whose solutions converge to $E_{\mathrm{sep}}$.  The
simplest program, which applies for bipartite systems with Hilbert
space $\mathcal{H}_A \otimes \mathcal{H}_B$, is
\begin{align}
  \label{pptgap}
  \max &\quad \epsilon \,, \nonumber \\
  \text{subject to} \quad  H - \epsilon I  &= P + Q^{T_A} \,, \nonumber \\
  P &\geq 0 \,, \nonumber \\
  Q &\geq  0 \,, 
\end{align}   
where $T_A$ denotes the partial transpose over system $A$.  Let $d_i =
\dim(\mathcal{H}_i)$, $i = A,B$.  When $d_A = 2$ and $d_B = 2$ or $3$
the maximum $\epsilon$ obtained from this program corresponds to the
minimum separable energy, $E_{\mathrm{sep}}$.  The optimum value
$\epsilon^*$ of the semidefinite program gives an entanglement witness
$Z_{EW} = H - \epsilon^* I$, and a lower bound on the entanglement gap
equal to the largest magnitude negative eigenvalue of $Z_{EW}$.  The
entanglement witness produced by (\ref{pptgap}) is referred to as
\emph{decomposable} because it can be written $Z_{EW}=H - \epsilon^* I
= P + Q^{T_A}$ for $P \geq 0$, $Q \geq 0$, and can only detect
entangled states with non-positive partial transpose.

If the subsystems are of higher dimension, it is possible for an
entangled state to have a positive partial transpose.  Such states are
\emph{bound entangled} \cite{horodecki2001a}, and the semidefinite
program (\ref{pptgap}) only finds the gap between the ground-state
energy and the minimum-energy positive partial transpose state.  This
solution provides a lower bound on $E_{\mathrm{sep}}$, and it is
possible to devise a nested sequence of programs that provide
increasingly tighter bounds~\cite{Doherty2004a}.

As all entanglement witnesses may be viewed as Hamiltonians with
entangled low energy states, one way of producing bound entangled
states suggests itself: as thermal states.  An example of a
Hamiltonian for which there are bound entangled states which achieve
lower energy than any separable state may be derived from the Choi
form, as described in \cite{Doherty2004a,horodecki2001a}.  This
Hamiltonian, which acts on the minimal-dimension system on which bound
entangled states exist, i.e.\ $\dim(\mathcal{H}_A) =
\dim(\mathcal{H}_B) = 3$, is
\begin{multline}
  H = 2 (\qmproj{00} + \qmproj{11} + \qmproj{22})\\ + \qmproj{02}
   + \qmproj{10}+\qmproj{21} - 3 \qmproj{\psi_+} \,,
\end{multline}
where $\ket{\psi_+} = \frac{1}{\sqrt{3}} \sum_{i=0}^2 \ket{ii}$.  The
ground-state energy of this Hamiltonian is $-1$, the minimum
separable energy is 0 and there are bound entangled states with energy
as low as $(3-2\sqrt{3})/3 \simeq -0.1547$.  Although
$E_{\mathrm{sep}}=0$ the semidefinite program (\ref{pptgap}) would
return $-0.1547$ for this Hamiltonian, the energy of the minimum
energy positive partial transpose state.  Implementing higher order
programs as per \cite{Doherty2004a} would give more and more accurate
estimates of the true minimum separable energy, $E_{\mathrm{sep}} =
0$.  Furthermore there is a small range of temperatures, $1.256 \lesssim
k_B T \lesssim 1.271$, over which the thermal state has energy less
than zero, so it is certainly entangled, but has positive partial
transpose.  Over this range of temperatures the Hamiltonian witnesses
the bound entanglement of the thermal state.

Examples of Hamiltonians where \emph{all} low-energy states are bound
entangled may be constructed from \emph{unextendable product bases}
\cite{bennett1999a}: a set of product states for which the orthogonal
complement contains no product states.  To construct the Hamiltonian
we let the unextendable product basis span the excited-state manifold,
and its orthogonal complement the ground-state manifold.  In this
extreme example, all thermal states with energy within the
entanglement gap are bound entangled.

\section{Entanglement-Gap Temperature of Bipartite Systems}

\label{app: enttemp}

In this Appendix, we investigate the entanglement-gap temperature of
bipartite Hamiltonians.  For this purpose, we define a
\emph{completely entangled} subspace of a multipartite Hilbert space
as one that contains no separable states.  The antisymmetric subspace
of two systems is an example.  One might wonder whether it is possible
to find a Hamiltonian with a completely entangled ground-state
manifold that is larger than the antisymmetric subspace so as to
achieve a higher entanglement-gap temperature than the symmetric
projector (\ref{eq:SymProj}).  In \cite{P} the maximum dimension of a
completely entangled subspace of many parties was investigated: for
two $d-$dimensional systems a basis was given for a completely
entangled subspace of maximum possible dimension $d^2-2d+1$.  This
subspace contains the antisymmetric subspace.

A natural candidate for a Hamiltonian with a high entanglement-gap
temperature is thus the Hamiltonian with such a subspace at energy
zero and its orthogonal complement at the highest energy.  To find its
entanglement gap we could, in principle, use a sequence of
semidefinite programs as described in Appendix~\ref{app: semidefentgap}.
However, as the dimension increases we need to implement increasingly
higher order tests to ensure convergence and computer memory
requirements become prohibitive.  These programs always return a
\emph{lower bound} on the entanglement gap.  Alternatively, we can
bound the gap from above by choosing random pure product states
\footnote{To create random pure quantum states we draw the components
  from $\mathcal{N}(0,1)$, i.e.\ a normal distribution with mean zero
  and variance one, and normalize the state.  This sampling is
  equivalent to choosing states according to the Haar measure (see for
  example Appendices A and B of Ref.~\cite{BHLSW2003a}).} and
evaluating their energies.  The lowest energy of a large number of
trial states provides an \emph{upper bound} on the entanglement gap
and thus on the entanglement-gap temperature.
     
\begin{figure}[htb]
  \includegraphics[width=80mm]{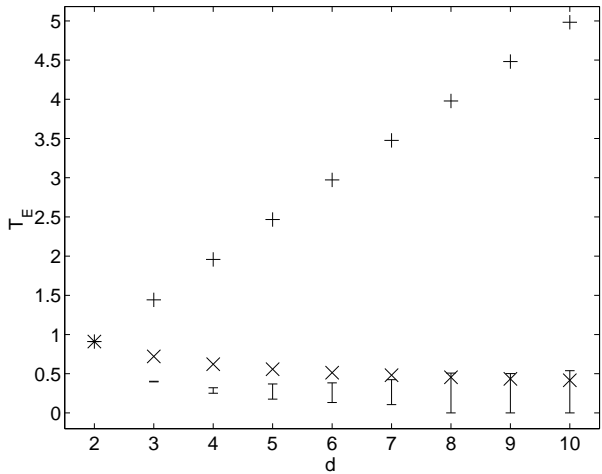}
\caption{Comparison of entanglement-gap temperature as a function of
  dimension of the subsystems for the three bipartite Hamiltonians:
  crosses correspond to $H_{\rm me} = I - \qmproj{\phi_d}$, pluses to
  $H_S =\Pi_S$ and bounding bars to $H_{\mathrm{ces}}
  = I - \Pi_{\mathrm{ces}}$.}
\label{fig: enttemp}
\end{figure}

Figure \ref{fig: enttemp} compares the behavior of the entanglement
gap temperature as a function of $d$ for the three Hamiltonians
considered above, $H_{\rm me} = I - \qmproj{\phi_d}$, $H_S = \Pi_S$
and $H_{\mathrm{ces}} = I - \Pi_{\mathrm{ces}}$, where
$\Pi_{\mathrm{ces}}$ is the projector onto the completely entangled
subspace of maximum dimension.  We see that the entanglement-gap
temperature of $H_{\mathrm{ces}} = I - \Pi_{\mathrm{ces}}$ is
generally comparable to that of $H_{\rm me} = I - \qmproj{\phi_d}$.
This result is due to the fact that the entanglement gap for
$H_{\mathrm{ces}}$ is quite small, thus resulting in low entanglement
gap temperature despite the large ground-state degeneracy.

Another method for constructing completely entangled subspaces is as
the orthogonal complement of \emph{unextendable product bases}
\cite{bennett1999a}.  We have constructed Hamiltonians with completely
entangled ground-state manifolds from a number of known unextendable
product bases and have always found entanglement-gap temperatures
significantly lower than that of the symmetric-projector Hamiltonian.
 
We thus have good evidence that the symmetric-projector Hamiltonian has the
highest entanglement-gap temperature of Hamiltonians with all energy
eigenvalues either zero or one.  The completely general case where
there can be intermediate energies as well is beyond the scope of this
work.

\section{Maximum Entanglement-Gap Temperature for Two Qubits}

\label{sec: maxenttempmegs}

In this Appendix we investigate the entanglement temperature of two
qubit systems, and provide evidence that the Heisenberg
antiferromagnetic Hamiltonian has the highest scaled entanglement-gap
temperature.  We scale all two-qubit Hamiltonians so that the
ground-state energy is zero, the maximum energy is one, and there are
two intermediate energies, $0 \leq E_1 \leq E_2 \leq 1$.  The
antiferromagnet has the singlet at energy zero and all triplet states
at energy one; its scaled entanglement gap and entanglement-gap
temperature are $g_E = 1/2$ and $t_E = 1/\log_e (3)$.

We present two lemmas leading to a theorem that any Hamiltonian with a
maximally entangled ground state has an entanglement-gap temperature
lower than that of the Heisenberg antiferromagnet.

\begin{lem}
\label{lem: thermthresh}
Let $H$ and $H'$ be two multipartite Hamiltonians with entanglement
gap temperatures $T_E$ and $T'_E$, respectively.  If there is a
separable state, $\rho_{\mathrm{sep}}$ such that
\begin{equation}
  \tr[H' \rho_{\mathrm{sep}}]  \leq U'(T_E)\,,
\end{equation}  
where $U'(T)$ is the thermal energy of $H'$, then $T'_E \leq T_E$.
\end{lem}

\noindent
\textbf{Proof:} $\tr[H' \rho_{\mathrm{sep}}]$ is an upper bound on
$E'_{\mathrm{sep}} = \min_{\rho_{\mathrm{sep}} \in \mathcal{S}} \tr[H'
\rho_{\mathrm{sep}}]$.  By definition $U'(T'_E) = E'_{\mathrm{sep}}$,
so the result follows from the fact that $U'(T)$ is a monotonically
increasing function of $T$. \hfill$\Box$\medskip

\begin{lem}
  \label{lem: E1leq14}
  Any Hamiltonian, $H'$, with $E_1 \leq 1/4$ has an entanglement-gap
  temperature less than that of the Heisenberg antiferromagnet.
\end{lem}

\noindent
\textbf{Proof:} We use the fact that, for two qubits, all
two-dimensional subspaces contains a separable state~\cite{P}.  Thus,
there must be a separable state in the subspace spanned by $\ket{E_0}$
and $\ket{E_1}$, and this separable state must have energy less than
or equal to $E_1$.

We now apply Lemma \ref{lem: thermthresh} with this separable state
$\rho_{\mathrm{sep}}$.  Because $E_1$ is the lower of the two
intermediate energies, the Hamiltonian, $H''$, with the same
eigenstates and eigenenergies as $H'$, except that $E_2 = E_1$ will
certainly have a lower thermal energy at any particular temperature
than $H'$, $U''(T) \leq U'(T),\ \forall\ T$.  The thermal energy
$U''(T)$ is easily calculated; with it, we find a value of $E_1$ that
satisfies the condition
\begin{equation}
  E_1 \leq U''(T_E = 1/\log_e(3)) \quad
  \Rightarrow \quad E_1 \leq 1/4 \,.
\end{equation}
Thus, if $E_1 \leq 1/4$ then $\tr[H' \rho_{\mathrm{sep}}] \leq U''(T_E
= 1/\log_e(3)) \leq U'(T_E = 1/\log_e(3))$, so $H'$ has a lower
entanglement-gap temperature than the Heisenberg antiferromagnet, as
required. \hfill$\Box$\medskip
  
\begin{thm}
  Any Hamiltonian, $H'$, with a maximally entangled ground state has
  an entanglement-gap temperature less than that of the Heisenberg
  antiferromagnet.
\end{thm}

\noindent
\textbf{Proof:} Given a Hamiltonian with a maximally entangled ground
state we can use local unitaries to transform to a Hamiltonian with
the singlet as its ground state $\ket{E_0} = (\ket{0} \ket{1} -
\ket{1} \ket{0})/\sqrt{2}$.  By Lemma \ref{lem: locunitary} and the
invariance of the spectrum under any unitary, this Hamiltonian has the
same entanglement-gap temperature.  The excited eigenstates for this
Hamiltonian all lie in the symmetric (triplet) subspace.  We express
the excited states in their Schmidt decompositions as $\ket{E_i} =
\lambda_i \ket{0_i} \ket{0_i} + \sqrt{1-\lambda_i^2}\ket{1_i}
\ket{1_i}$, where $i = 1,2,3$.

We present two separable states, one of which has energy less
than the threshold for any Hamiltonian.  The first is
$\rho_{\mathrm{sep}} = \qmproj{A} \otimes \qmproj{B}$ where
\begin{equation}
  \ket{A} = (\ket{0_1}+\ket{1_1})/\sqrt{2}\,, \quad
  \ket{B} = (\ket{0_1}-\ket{1_1})/\sqrt{2}\,.
\end{equation}
The energy of this state is at most $\tr[H' \rho_{\mathrm{sep}}] =
(E_1+1)/4$.  For a given $E_1$ this energy will be less than $U'(T_E)$
for $E_2$ greater than a certain lower bound, $E_{2}^{lb}$.
$E_2^{lb}(E_1)$ is defined implicitly by
\begin{equation}
  \label{eq: E2lb} 
  \tr[H' \rho_{\mathrm{sep}}]  = U'(T_E = 1/\log_e(3))\,.
\end{equation}
This equation is transcendental and so it is not possible to find an
explicit functional form for $E_2^{lb}(E_1)$.

The second low-energy separable state that we consider is
$\rho_{\mathrm{sep}} = \qmproj{A} \otimes \qmproj{B}$ where $\ket{A} =
\ket{0_3}$ and $\ket{B} = \ket{1_3}$.  The energy of this state is at
most $\tr[H' \rho_{\mathrm{sep}}] = E_2/2$.  This energy will be less
than $U'(T_E = 1/\log_e(3))$ for $E_2$ less than a maximum value,
$E_2^{ub}(E_1)$, defined by
\begin{equation}
  \label{eq: E2ub}
  E_2/2 = U'(T_E = 1/\log_e(3))\,.
\end{equation}
We numerically solve the two equations, (\ref{eq: E2lb}) and (\ref{eq:
  E2ub}), for $E_2^{lb}(E_1)$ and $E_2^{ub}(E_1)$, respectively.  From
Lemma~\ref{lem: thermthresh} it is only possible that $T_E' > T_E =
1/\log_e(3)$ if $E_1 > 1/4$.  However, it is straightforward to
calculate numerically $E_2^{lb}(E_1) \leq E_2^{ub}(E_1)$ in this
region, so that for any $(E_1,E_2)$, there is a separable state with
energy less than $U'(T_E = 1/\log_e(3))$.  Lemma~\ref{lem: E1leq14}
requires that $T'_E \leq T_E$, so the Heisenberg antiferromagnet has
the highest entanglement-gap temperature of any bipartite Hamiltonian
with a maximally entangled ground state.  \hfill$\Box$\medskip

A generic two-qubit Hamiltonian has a non-maximally entangled ground
state, so it is still possible that such a Hamiltonian possesses a
higher entanglement-gap temperature than the Heisenberg
antiferromagnet.  To provide numerical evidence that such a
Hamiltonian does not exist, we generated random Hamiltonians by
drawing the two intermediate energy levels from a uniform
distribution.  Because no bound entangled states exist for two qubits,
the semidefinite program of Appendix~\ref{app: semidefentgap} produces
the entanglement gap.  We then calculated the entanglement gap
temperature numerically.  We generated $10^8$ random Hamiltonians and
calculated their entanglement-gap temperature in this way.  None were
found to have an entanglement-gap temperature higher than that of the
Heisenberg antiferromagnet, providing strong evidence that it
possesses the highest possible entanglement-gap temperature.

\section{Transverse Field $XY$ Model}

\label{app: TFXY}

The transverse field $XY$ model is defined by the coupling Hamiltonian
Eq.~(\ref{eq: XYhamil}).  To find the minimum-energy separable state
$\ket{A} \ket{B}$ we parameterize the two factors as
\begin{equation}
  \label{eq: ABfactors}
  \ket{j} = \cos \theta_j \ket{\uparrow} + e^{i \phi_j} \sin
  \theta_j \ket{\downarrow}\,, \quad j=A,B\,,
\end{equation}
where $0 \leq \theta_j \leq \pi/2$, $0 \leq \phi_j <
2 \pi$.  We then calculate the energy of the product state $\ket{A}
\ket{B}$ as a function of the four parameters:
\begin{multline}
  \bra{A} \bra{B} H^{XY}_{AB} \ket{A} \ket{B} = \tfrac{\lambda}{2}
 (\cos 2 \theta_A + \cos 2 \theta_B) \\ 
 + \left( \tfrac{1+\gamma}{2} \right) \cos \phi_A \sin 2 \theta_A \cos
  \phi_B \sin 2 \theta_B \\
  + \left( \tfrac{1-\gamma}{2} \right) \sin \phi_A \sin 2 \theta_A \sin
  \phi_B \sin 2 \theta_B \,,
\end{multline}
and optimize over this space to find the lowest energy separable
state.  The result is:
\begin{equation}
  \label{eq: XYlowsep}
  \ket{j} = \begin{cases}
  \sqrt{\frac{1+ \gamma + \lambda}{2(1+\gamma)}} \ket{\uparrow} \pm
  \sqrt{\frac{1+ \gamma -  \lambda}{2(1+\gamma)}} \ket{\downarrow}, &
  \lambda \leq 1 + \gamma,  \\
  \ket{\downarrow}, & \lambda \geq 1 + \gamma
  \end{cases}
\end{equation}
where the $\pm$ corresponds to $j=A,B$, with energy
\begin{equation}
\label{eq: EsepXY} E_{\mathrm{sep}}^{XY} = \begin{cases}
  -\frac{(1+\gamma)^2+ \lambda^2}{2 (1 + \gamma)}\,,& \lambda \leq 1
  + \gamma \\
  - \lambda\,,& \lambda \geq 1 + \gamma \,. \end{cases}
\end{equation}
Incidentally, by calculating the spectrum of $H^{XY}_{AB}$ we can
identify a curve, $ \lambda^2 + \gamma^2 = 1$ on which there is a
separable state in the degenerate ground-state manifold.  The
entanglement gap is therefore zero on this curve, and this result
remains true for the $XY$ model on an arbitrary bipartite lattice
(with the appropriate magnetic field in the coupling Hamiltonian).

%\bibliographystyle{apsrev} 
%\bibliography{Cav13_mrd}

\end{document}